\documentclass[twocolumn]{revtex4}
\usepackage{graphicx}
\usepackage{latexsym}

\usepackage{color}

\def\be{\begin{equation}}
\def\ee{\end{equation}}
\def\bea{\begin{eqnarray}}
\def\eea{\end{eqnarray}}

\begin{document}

\title{Evidences for bouncing evolution before inflation in cosmological surveys}

\author{Jie Liu${}^{a}$\footnote{Email: liujie@ihep.ac.cn}}
\author{Yi-Fu Cai${}^{b}$\footnote{Email: ycai21@asu.edu}}
\author{Hong Li${}^{a,c}$\footnote{Email: hongli@ihep.ac.cn}}

\affiliation{${}^a$ Institute of High Energy Physics, Chinese
Academy of Science, P.O.Box 918-4, Beijing 100049, P.R.China}

\affiliation{${}^b$ Department of Physics, and School of Earth and
Space Exploration, and Beyond Center, Arizona State University,
Tempe, AZ, 85287-1504, USA}

\affiliation{${}^c$ Theoretical Physics Center for Science
Facilities (TPCSF), Chinese Academy of Science, P.R.China}

%\date{\today}

\begin{abstract}
Inflationary cosmology with a preceding nonsingular bounce can
lead to changes on the primordial density fluctuations. One
significant prediction is that the amplitude of the power spectrum
may undergo a jump at a critical scale. In this Letter we propose
a phenomenological parametrization of the primordial power
spectrum in this scenario and confront the jump feature with
latest cosmological data. Performing a global fitting, we utilize
this possibility to derive a novel method for constraining bounce
parameters via cosmological measurements. Combining the CMB, LSS
and SNIa data, our result interestingly reveals that a nonsingular
bounce, if exists, should be a fast bounce which happens at a very
high energy scale, as we get an upper limit on the bounce
parameters.
\end{abstract}

%\pacs{98.80.Es, 11.30.Cp, 11.30.Er, 98.80.Cq}

\maketitle

%Introduction==========================================================
\section{Introduction}\label{Int}

Astronomical observations favor an adiabatic and nearly
scale-invariant power spectrum of primordial perturbation, which
can be realized in an inflation model by requiring a cosmological
scalar field slowly rolling along the plateau of its potential.
Based on this, it is usually assumed that the primordial power
spectrum of cosmological perturbation is a fixed power function of
the comoving wavenumber $k$ over the range of observable scales in
the detailed methods of data fitting. However, as advocated by
physics of the very early universe, namely the trans-Planckian
physics\cite{Brandenberger:1999sw, Martin:2000xs}, bounce
cosmologies\cite{Mukhanov:1991zn, Cai:2007qw} and so on, it is
possible that the primordial spectrum shows local features which
cannot be described by the usual
parameterizations\cite{Cai:2008qb, Piao:2003zm}.

As well-known, an inflation model suffers from the problem of
initial singularity and thus one cannot use the effective field
approach to describe the universe at the beginning of its
evolution\cite{Borde:1993xh}. This problem can be circumvented in
the framework of bounce cosmology, where the big bang singularity
is replaced by a nonsingular bounce. By virtue of the effective
field description, one can realize a bounce through an effective
violation of certain energy condition, and thus obtain an
inflationary scenario with a preceding bounce\cite{Cai:2008qb}. In
this scenario, however, the cosmological evolution is usual
asymmetric with respect to the bounce point since of back-reaction
of primordial perturbation and
radiation\cite{Brandenberger:2009rs}. Therefore, there are two
plausible scenarios for primordial perturbations to exit the
Hubble radius, with one being during a phase of matter-dominated
contraction\cite{Wands:1998yp, Starobinsky:1979ty} and the other
being inflation after the bounce. If the initial perturbation
originate from a Bunch-Davies vacuum and the contracting phase
connects to an expanding one via a nonsingular bounce, the spectra
produced in both stages are almost scale-invariant without changes
as long as the scales we are interested are much larger than the
duration of the bounce. However, they are different in their
amplitudes, and thus their combination yields a nearly
scale-invariant primordial power spectrum with a jump feature at a
critical scale.

In the current Letter, we phenomenologically propose a
parametrization of primordial power spectrum with a jump feature
motivated by the bounce inflation scenario and study the
constraints on this feature from current astronomical
observations. Compared to the standard inflationary paradigm, this
model involves two additional parameters, $k_B$ characterizing the
comoving wavelength of the universe at the bounce point, and $T$
describing the slope of the jump in the power spectrum and thus
corresponding to the measurement of the duration of the bouncing
phase. In the literature, the technique of Markov Chain Monte
Carlo (MCMC) global analysis has been widely generalized to
constrain non-standard inflationary cases, namely non-canonical
inflation models\cite{Peiris:2007gz, Hall:2007qw}, trans-Planckian
physics\cite{Easther:2004vq, Spergel:2006hy}, and so on. It has
proven to be very powerful to probe the parameter space beyond the
standard inflationary paradigm. As a consequence, we employ the
MCMC technique to do a global fitting to constrain parameters of
bounce inflation and give a comparison to current data.

The letter is organized as follows. In Section II we briefly
review the cosmological perturbations generated in a bounce model
with a matter contraction connecting to an inflation, and present
a smoothed parametrization of the primordial power spectrum of
this model with a jump feature. In Section III we perform a
numerical calculation and compare the result with the CMB and LSS
data. We give our numerical results in Section IV before
concluding with a discussion in Section V.

%Method and Current Observations=======================================
\section{Formalism in a Bounce-Inflation Scenario}

We begin with a brief discussion of the cosmological evolution of
primordial perturbation in the framework of a flat FRW Universe. A
standard process of generating primordial power spectrum suggests
that, cosmological fluctuations should initially emerge inside a
Hubble radius, then leave it in the primordial epoch, and
finally reenter at late times. In usual, this process can be
realized by stretching the physical wavelength $\frac{a}{k}$
longer than the Hubble radius $1/H$ as in inflation; while an
alternative is to suppress the comoving Hubble radius $1/aH$
shorter than the comoving wavelength $1/k$ as in matter bounce.

One often uses a gauge-invariant variable $\zeta$, the curvature
fluctuation in comoving coordinates, to characterize the
cosmological inhomogeneities. It is associated with a canonical
variable $v=z\zeta$, where $z\equiv \sqrt{2\epsilon}a$ with
$\epsilon\equiv -\dot H/H^2$. The equation of motion for the
Fourier mode $v_k(\eta)$ in the context of standard Einstein
gravity is given by\footnote{The generic equation of motion for
cosmological fluctuations involves a sound speed parameter in
front of the gradient term\cite{Mukhanov:1990me}.}
\begin{eqnarray}\label{eom}
 v_k''+(k^2-\frac{z''}{z})v_k=0~,
\end{eqnarray}
where the prime denotes the derivative with respect to the
comoving time $\eta\equiv \int dt/a$.

To perform a specific analysis, one may take the scale factor as
$a(t)=a_B(\frac{t}{t_B})^{1/\epsilon}$, where the subscript ``$B$"
denotes any reference time which will be referred as the bouncing
point later. For a constant background equation-of-state (EoS)
$w$, one obtains
\begin{eqnarray}\label{mass}
 \frac{z''}{z}=\frac{\nu^2-\frac{1}{4}}{\eta^2}~,~
 {\rm with}~~\nu=\pm\frac{\epsilon-3}{2(\epsilon-1)}~.
\end{eqnarray}

We assume the cosmological perturbations originate from vacuum
fluctuations, which suggests
\begin{eqnarray}\label{inicond}
 v_k^{i}\simeq\frac{1}{\sqrt{2k}}e^{-i\int^\eta kd\tilde\eta}~,
\end{eqnarray}
when fluctuations are born with $|k\eta|\gg1$. This is consistent
with the asymptotic solution to Eq.(\ref{eom}) when the last term
$\frac{z''}{z}$ is negligible. Therefore, the mechanism of
generating primordial perturbations requires that the absolute
value of the comoving time is enough large which can only be
achieved in a contracting or an inflationary setup. Another
asymptotic solution to Eq.(\ref{eom}) can be derived by virtue of
the mathematic property of the Bessel function,
\begin{eqnarray}\label{sollead}
 v_k \sim \eta^{\frac{1}{2}} \bigg[ c(k)\eta^{-|\nu|} \bigg]~,
\end{eqnarray}
in the super-Hubble regime, which implies $|k\eta|\ll1$.

Now we match the two asymptotic solutions (\ref{inicond}) and
(\ref{sollead}) at the moment of Hubble crossing $|k\eta|\sim1$,
and thus determine the form of $v_k$ on super-Hubble scale,
\begin{eqnarray}\label{solution}
 v_k(\eta) \simeq \frac{1}{\sqrt{2k}} (k\eta)^{\frac{1}{2}-|\nu|}~.
\end{eqnarray}
From the definition of the power spectrum, one learns that
$\zeta\sim k^{3/2}|v_k|$ can only be scale-invariant when
$|\nu|=3/2$. As a consequence, the primordial fluctuations are
nearly scale-invariant in both the matter contraction and
inflationary scenarios. However, $\epsilon$ takes the value
${3}/{2}$ in the matter contraction but becomes very small during
inflation. The amplitude of the primordial spectrum would undergo
a jump around the scale comparable to the bounce scale. A detailed
calculation gives the expression of the primordial power spectrum
for a model of bounce inflation as follows
\begin{eqnarray}\label{spectrum}
 P_{\zeta} = \left\{ \begin{array}{c}
               \frac{H^2}{48\pi^2}~,~~k<k_B \\
               \\
               \frac{H^2}{8\pi^2\epsilon}~,~k\geq{k}_B~,
\end{array} \right.
\end{eqnarray}
where the parameter $k_B$ characterizes the comoving wavelength of
the universe at the bouncing point. Note that, when the
perturbation passes through the bouncing phase, its positive and
negative frequency modes could be mixed at the transfer surface
and thus bring an oscillating signal around the bounce
scale\cite{Cai:2008qb}. However, this signal strongly depends on
extra parameters introduced in specific models and is not quite
sensitive to current observations. In this Letter we have smoothed
this signal and study its average effect directly.

For phenomenological considerations, we would like to parameterize
the above spectrum by assuming a form
\begin{eqnarray}\label{parametrize_p}
 P_{\zeta}=P_m+\frac{P_{inf}-P_m}{2}\bigg\{1+\tanh[(k-k_B)T]\bigg\}~,
\end{eqnarray}
where $P_{inf}=\frac{H^2}{8\pi^2\epsilon}$ and $P_m$ is the
spectrum of curvature perturbation before the bounce.
Theoretically, $P_{m}=\frac{H^2}{48\pi^2}$, which is relevant to
primordial tensor fluctuations, should be less than $1\%$ of
$P_{inf}$ due to the upper bound of tensor-to-scalar ratio.
% $r<0.49$ at $2\sigma$.
The power spectrum after the bounce $P_{inf}$ is parameterized as
the power low format via $P_{inf}=A_s k^{n_s-1}$, in which $A_s$
and $n_s$ are the amplitude and the spectral index
correspondingly. $k_B$ is a comoving wavenumber relevant to the
bounce scale and denotes the occurrence moment of the jump feature
in the power spectrum (in unit of $h~{\rm Mpc^{-1}}$). The
parameter $T$ characterizes the slope of the jump feature (in unit
of ${\rm Mpc}$) and thus the duration of the bouncing phase. In
the bounce inflation scenario, we have $k_BT\sim{H}\Delta{t}_B$.
For this description, the phenomenology of the scale-invariant
primordial power spectrum will be recovered when $k-k_B> 0$ and
$T\rightarrow\infty$. In the following we study observational
constraints from current data on the bounce parameters introduced
in Eq.(\ref{parametrize_p}).

%Numerical method===============================================================
\section{Constraints from observational Data}
\subsection{signal in CMB and LSS}

\begin{figure}[htbp]
\begin{center}
\includegraphics[scale=0.9]{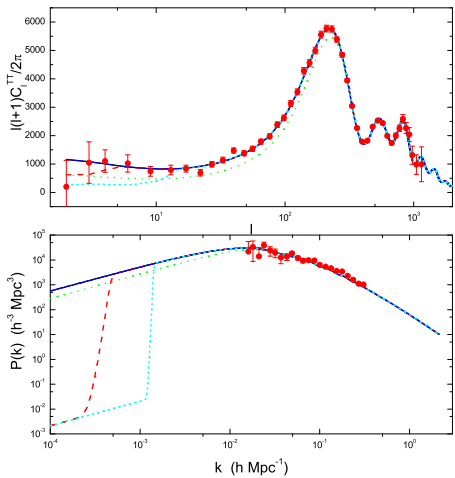}
\caption{Signatures of bounce models in the temperature power
spectrum (top) and the matter power spectrum (bottom). We compare
a $\Lambda$CDM model (black solid line) with four bounce models
with different parameters as introduced in the main context. The
red dots are the observational data points from CMB (top) and LSS
(bottom).\label{Fig:ClTT}}
\end{center}
\end{figure}

To test our model we first consider the current astronomical
observations from the Wilkinson Microwave Anisotropy Probe 7-year
data (WMAP7)\cite{Komatsu:2010fb} and the Sloan Digital Sky Survey
(SDSS)\cite{Tegmark:2003uf}. Similar to the trans-Planckian
physics, the effect brought by a bounce is most sensitive to the
modes of primordial perturbation exiting the Hubble radius at
earliest time. Therefore, the bounce induced jump feature of the
primordial power spectrum will imprint its effects on the CMB at
very large length scales, and correspondingly depress the CMB
anisotropies at large angular scales. Fig. \ref{Fig:ClTT}
illustrates this impact on the CMB temperature power spectrum. We
compare a $\Lambda$CDM power spectrum (black solid line) with
$\omega_b=0.023$, $\omega_c=0.11$, $h=0.71$, $\tau=0.09$,
$A_s=2.2\times10^{-9}$, $n_s=0.97$ and four bounce models with
different power spectra, namely $P_m=5\times10^{-12}$, $T=2000$,
$k_B=3\times10^{-4}$ (red dashed line), $P_m=10^{-11}$, $T=4$,
$k_B=8\times10^{-6}$ (green dotted line), $P_m=10^{-10}$,
$T=2000$, $k_B=8\times10^{-6}$ (blue dash dotted line) and
$P_m=10^{-11}$, $T=2000$, $k_B=10^{-3}$ (cyan short dashed line).

At large angular scale, the initial condition is given by $P_m$
rather than $P_{inf}$, so the reduction in the amplitude of the
power spectrum happens beyond the bounce scale where $k<k_B$. This
effect can be seen by comparing the black solid line with the blue
dashed line. Moreover, the smaller value of $P_m$ we choose, the
effect of this reduction is more obvious at large scale. The
parameter $k_B$ is determined by the energy scale of the bounce,
and thus the minimal size of the universe in this scenario. From
Fig. \ref{Fig:ClTT}, we can find that the amplitude of primordial
power spectrum would obtain an increase around the scale of $k_B$.
In numerical computation, we illustrate this effect by choosing
$k_B=3\times10^{-4}$ and $k_B=10^{-3}$ respectively. Additionally,
the parameter $T$ influences the shape of the temperature power
spectrum obviously. If $T$ is large enough, the jump feature of
the power spectrum is shown to be very apparent. However, if we
choose a small value of $T$, the whole shape of the power spectrum
becomes smooth relatively and a quite sizable regime of this
spectrum could be suppressed manifestly. It can be seen by the
green dotted line with $T=4$, which is too small to accommodate
the WMAP data, and this feature could lead to a constraint on $T$
as will be analyzed later.

The jump feature in bounce inflation can also leave its signature
on the matter power spectrum on large scales as shown in the lower
panel of Fig. \ref{Fig:ClTT}. Since the signature of the bounce
effect occurs only at the edge of the observable regime of the
current data of LSS and accordingly we can only obtain a bound
instead of an accurate constraint. Our global analysis provides a
powerful approach towards a future detection of bounce cosmologies
making use of data from more accurate astronomical experiments.

\subsection{MCMC Likelihood Analysis}\label{Method}

The MCMC technique is widely applied to give multi-dimensional
parameter constraints from observational data. In the current
Letter, we employ MCMC to generate a random sample from the
posterior distribution ${\cal P}( \theta|x)$ of a set of
parameters $\theta$ given an event $x$ (for us, it is the total
data set that used), and obtain
\begin{eqnarray}\label{eq:bayes}
 {\cal P}(\mbox{\boldmath $\theta$}|x)
  = \frac{{\cal P}(x|\theta){\cal P}(\theta)}{\int {\cal P}(x|\theta){\cal P}(\theta)d\theta}~,
\end{eqnarray}
via Bayes' Theorem, where ${\cal P}(x|\theta)$ is the likelihood
of event $x$ given the model parameters $\theta$ and ${\cal
P}(\theta)$ is the prior probability distribution of obtaining a
model parameter value $\theta$. The simulated random observations
from the posterior distribution are of the likelihood
%%RB What do you mean by "of the likelihood surface?
surface, and from this sample, we can estimate the posterior
distribution of the parameter of interest.

For our implementation, we use a generalized version of the
CosmoMC package\cite{Lewis:2002ah}, in which the cosmological
parameters are: %%
\begin{eqnarray} \label{parameter}
 {\cal \theta} \, \equiv \, (\omega_b, \omega_c, \Theta_s, \tau, n_s, A_s, P_m, k_B, T)~,
\end{eqnarray}
where $\omega_{b}\equiv\Omega_{b}h^{2}$ and
$\omega_{c}\equiv\Omega_{c}h^{2}$, in which $\Omega_{b}$ and
$\Omega_{c}$ are the baryon and cold dark matter densities
relative to the critical density,  $\Theta_{s}$ is the ratio
(multiplied by 100) of the sound horizon to the angular diameter
distance at decoupling, and $\tau$ is the optical depth to
re-ionization. The remaining parameters are related to the
primordial power spectrum given by Eq.(\ref{parametrize_p}). For
simplicity, we assume a purely adiabatic spectrum of fluctuations
and a flat universe with a cosmological constant whose EoS is
$w=-1$ at the initial moment.

In our analysis, we have included the WMAP7 temperature and
polarization power spectra with the routine for computing the
likelihood supplied by the WMAP team, the matter power spectrum
from the ``Luminous Red Galaxies" sample from the
SDSS\cite{Tegmark:2003uf} as well as the SNIa ``Union" compilation
(307 sample)\cite{Kowalski:2008ez}. In calculating the likelihood
from SNIa we have marginalized over the ``nuisance
parameter"\cite{DiPietro:2002cz}. Furthermore, we make use of the
Hubble Space Telescope (HST) measurement of the Hubble parameter
$H_{0}\equiv 100h~km~s^{-1}{\rm Mpc^{-1}}$ by applying a Gaussian
likelihood function centered around $h=0.742$ with standard
deviation $\sigma=0.038$\cite{Riess:2009pu}, and take the total
likelihood to be the products of the separate likelihoods of CMB,
SNIa and LSS. Alternatively defining $\chi^2 = -2 \log {\bf
\cal{L}}$, we get
\begin{eqnarray}
 \chi^2_{total} = \chi^2_{CMB}+ \chi^2_{SNIa}+\chi^2_{LSS}~.
\end{eqnarray}

%Results===============================================================
\section{Combined constraints}

From numerical computations, we find that the present data are far
from putting explicit constraint on bounce parameters on current
observationally accessible scales. However, the situation will
improve greatly with the appearance of more and more accurate CMB
experiments in the near future. With the current data we can
obtain upper bounds on $P_m$, $k_B$ and a constraint on $T$. From
the numerical results, we find that $P_m<7.03\times10^{-11}$,
$k_B<2.44\times10^{-4}$, and $2.63\times10^{2}<T<7.98\times10^{5}$
at $2\sigma ~ C. L.$. The result also shows that a bounce model
gives a slightly better (lower) $\chi^2$ for the dataset than the
$\Lambda$CDM model by $1.4$. Accordingly one may conclude that a
bounce inflation model is quite efficient to explain cosmological
observations at large scale.

\begin{figure}[htbp]
\begin{center}
\includegraphics[scale=0.23]{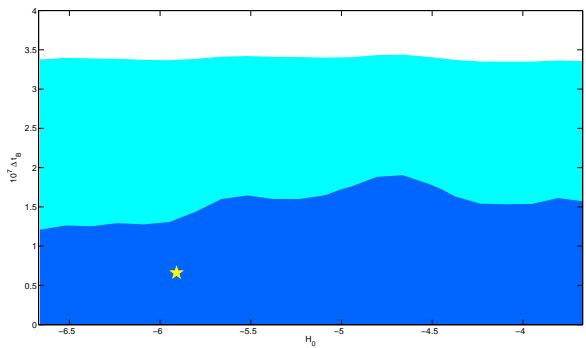}
\caption{A contour plot of $H$-$\Delta{t}_B$ from combined
constraints on bounce parameters $P_m$, $k_B$ and $T$ from CMB $+$
LSS $+$ SNIa data in the frame of bounce inflation.
\label{fig:kBT}}
\end{center}
\end{figure}

Finally, we calculated the combined constraints on the
inflationary Hubble parameter $H$ and the duration of bouncing
phase $\Delta{t}_B$, present the contour plot at $1\sigma$ and
$2\sigma$ respectively in Fig. \ref{fig:kBT}. The yellow star
represents a non-vanishing best-fit point. We also studied the
combined constraints on the inflationary spectral index $n_s$ and
its amplitude $A_s$ using global fitting. Although the derivation
is very limit, we find that a nonsingular bounce before inflation
leads to a larger red tilt for the spectral index and to more
power for primordial fluctuations compared with the $\Lambda$CDM
model.

%Summary===============================================================
\section{Summary}\label{Sum}

Since a nonsingular bounce happened at an extremely high energy
scale, it can hardly be tested directly by experiments. To find
evidence for it, we need to know its observational consequences.
This issue has been discussed widely in the literature, and one
potential clue is to study primordial perturbations. For example,
in the context of the Pre-Big-Bang scenario\cite{Gasperini:1992em}
and in the cyclic/Ekpyrotic model\cite{Khoury:2001wf}, the
resulting cosmological perturbation was found to strongly depend
on the physics at the epoch of thermalization, and thus an
uncertainty for prediction is involved\cite{Brustein:1994kn,
Lyth:2001pf}. In this Letter, motivated by a combined scenario of
matter bounce and inflation, we introduce a parametrization of the
primordial power spectrum in a class of bounce cosmologies, and
introduce two parameters to characterize the relevant physics. By
performing a global analysis, we show that the present
observations are quite consistent with a featureless power
spectrum, and consequently put a strong constraint on bounce
cosmologies, which suggests a fast bounce model at very high
energy scale. Although far from being conclusive, our numerical
method can be extended to other bounce cosmologies, and thus is
relevant in light of forthcoming astronomical observations, such
as PLANCK.

Note added: while this Letter was being finalized, a paper
appeared pointing out that the scenario of bounce inflation could
be embedded into loop quantum cosmology\cite{Mielczarek:2010ga},
and the data analysis in that work shows an abnormal enhancement
at the first peak of CMB anisotropy. We find it more generic to
consider the bouncing physics in a model-independent way as
introduced in the current Letter.

%Acknowledgments=======================================================
\section*{Acknowledgements}

{\it Acknowledgments --} We acknowledge the use of the Legacy
Archive for Microwave Background Data Analysis (LAMBDA). We thank
Robert Brandenberger, Eiichiro Komatsu, Jun-Qing Xia and Xinmin
Zhang for helpful comments and discussion. This work is supported in
part by the National Natural Science Foundation of China under Grant
No. 10803001 and 973 program No. 2010CB833000 and the Youth
Foundation of the institute of high energy physics under Grant Nos.
H95461N.

%End===================================================================


\begin{thebibliography}{nn}

%====================Inflation=====

%\cite{Guth:1980zm}
%\bibitem{Guth:1980zm}
  %A.~H.~Guth,
  %``The Inflationary Universe: A Possible Solution To The Horizon And Flatness
  %Problems,''
  %Phys.\ Rev.\  D {\bf 23}, 347 (1981);
  %%CITATION = PHRVA,D23,347;%%
%
%\cite{Albrecht:1982wi}
%\bibitem{Albrecht:1982wi}
  %A.~Albrecht and P.~J.~Steinhardt,
  %``Cosmology For Grand Unified Theories With Radiatively Induced Symmetry
  %Breaking,''
  %Phys.\ Rev.\ Lett.\  {\bf 48}, 1220 (1982);
  %%CITATION = PRLTA,48,1220;%%
%
%\cite{Linde:1981mu}
%\bibitem{Linde:1981mu}
  %A.~D.~Linde,
  %``A New Inflationary Universe Scenario: A Possible Solution Of The Horizon,
  %Flatness, Homogeneity, Isotropy And Primordial Monopole Problems,''
  %Phys.\ Lett.\  B {\bf 108}, 389 (1982).
  %%CITATION = PHLTA,B108,389;%%

%====================Trans Planckian=====

%\cite{Brandenberger:1999sw}
\bibitem{Brandenberger:1999sw}
  R.~H.~Brandenberger,
  %``Inflationary cosmology: Progress and problems,''
  arXiv:hep-ph/9910410.
  %%CITATION = HEP-PH/9910410;%%

%\cite{Martin:2000xs}
\bibitem{Martin:2000xs}
  J.~Martin and R.~H.~Brandenberger,
  %``The trans-Planckian problem of inflationary cosmology,''
  Phys.\ Rev.\  D {\bf 63}, 123501 (2001);
  %[arXiv:hep-th/0005209].
  %%CITATION = PHRVA,D63,123501;%%
%
%\cite{Brandenberger:2000wr}
%\bibitem{Brandenberger:2000wr}
  %R.~H.~Brandenberger and J.~Martin,
  %``The robustness of inflation to changes in super-Planck-scale physics,''
  Mod.\ Phys.\ Lett.\  A {\bf 16}, 999 (2001).
  %[arXiv:astro-ph/0005432].
  %%CITATION = MPLAE,A16,999;%%

%====================Bounce cosmology=====

%\cite{Mukhanov:1991zn}
\bibitem{Mukhanov:1991zn}
  V.~F.~Mukhanov and R.~H.~Brandenberger,
  %``A Nonsingular universe,''
  Phys.\ Rev.\ Lett.\  {\bf 68}, 1969 (1992);
  %%CITATION = PRLTA,68,1969;%%
%
%\cite{Brandenberger:1993ef}
%\bibitem{Brandenberger:1993ef}
  R.~H.~Brandenberger, V.~F.~Mukhanov and A.~Sornborger,
  %``A Cosmological theory without singularities,''
  Phys.\ Rev.\  D {\bf 48}, 1629 (1993).
  %[arXiv:gr-qc/9303001].
  %%CITATION = PHRVA,D48,1629;%%

%\cite{Cai:2007qw}
\bibitem{Cai:2007qw}
  Y.~F.~Cai, T.~Qiu, Y.~S.~Piao, M.~Li and X.~Zhang,
  %``Bouncing Universe with Quintom Matter,''
  JHEP {\bf 0710}, 071 (2007);
  %[arXiv:0704.1090 [gr-qc]].
  %%CITATION = JHEPA,0710,071;%%
%
%\cite{Cai:2007zv}
%\bibitem{Cai:2007zv}
  Y.~F.~Cai, T.~Qiu, R.~Brandenberger, Y.~S.~Piao and X.~Zhang,
  %``On Perturbations of Quintom Bounce,''
  JCAP {\bf 0803}, 013 (2008).
  %[arXiv:0711.2187 [hep-th]].
  %%CITATION = JCAPA,0803,013;%%

%\cite{Novello:2008ra}
%\bibitem{Novello:2008ra}
  %for a review see:
  %M.~Novello and S.~E.~P.~Bergliaffa,
  %``Bouncing Cosmologies,''
  %Phys.\ Rept.\  {\bf 463}, 127 (2008).
  %[arXiv:0802.1634 [astro-ph]].
  %%CITATION = PRPLC,463,127;%%

%====================Bounce inflation=====

%\cite{Piao:2003zm}
\bibitem{Piao:2003zm}
  Y.~S.~Piao, B.~Feng and X.~m.~Zhang,
  %``Suppressing CMB quadrupole with a bounce from contracting phase to
  %inflation,''
  Phys.\ Rev.\  D {\bf 69}, 103520 (2004).
  %[arXiv:hep-th/0310206].
  %%CITATION = PHRVA,D69,103520;%%

%\cite{Cai:2008qb}
\bibitem{Cai:2008qb}
  Y.~F.~Cai {\it et al.},
  %T.~T.~Qiu, J.~Q.~Xia, H.~Li and X.~Zhang,
  %``A Model Of Inflationary Cosmology Without Singularity,''
  Phys.\ Rev.\  D {\bf 79}, 021303 (2009);
  %[arXiv:0808.0819 [astro-ph]].
  %%CITATION = PHRVA,D79,021303;%%
%
%\cite{Cai:2008ed}
%\bibitem{Cai:2008ed}
  Y.~F.~Cai and X.~Zhang,
  %``Evolution of Metric Perturbations in Quintom Bounce model,''
  JCAP {\bf 0906}, 003 (2009).
  %[arXiv:0808.2551 [astro-ph]].
  %%CITATION = JCAPA,0906,003;%%

%===============singularity==================

%\cite{Borde:1993xh}
\bibitem{Borde:1993xh}
  A.~Borde and A.~Vilenkin,
  %``Eternal Inflation And The Initial Singularity,''
  Phys.\ Rev.\ Lett.\  {\bf 72}, 3305 (1994).
  %[arXiv:gr-qc/9312022].
  %%CITATION = PRLTA,72,3305;%%

%===============asymmetirc bounce==========

%\cite{Brandenberger:2009rs}
\bibitem{Brandenberger:2009rs}
  R.~Brandenberger and X.~m.~Zhang,
  %``The Trans-Planckian Problem for Inflationary Cosmology Revisited,''
  arXiv:0903.2065 [hep-th];
  %%CITATION = ARXIV:0903.2065;%%
%
%\cite{Li:2009cu}
%\bibitem{Li:2009cu}
  H.~Li, J.~Q.~Xia, R.~Brandenberger and X.~Zhang,
  %``Constraints on Models with a Break in the Primordial Power Spectrum,''
  arXiv:0903.3725 [astro-ph.CO].
  %%CITATION = ARXIV:0903.3725;%%

%==================matter bounce===================

%\cite{Wands:1998yp}
\bibitem{Wands:1998yp}
  D.~Wands,
  %``Duality invariance of cosmological perturbation spectra,''
  Phys.\ Rev.\  D {\bf 60}, 023507 (1999);
  %[arXiv:gr-qc/9809062].
  %%CITATION = PHRVA,D60,023507;%%
%
%\cite{Finelli:2001sr}
%\bibitem{Finelli:2001sr}
  F.~Finelli and R.~Brandenberger,
  %``On the generation of a scale-invariant spectrum of adiabatic  fluctuations
  %in cosmological models with a contracting phase,''
  Phys.\ Rev.\  D {\bf 65}, 103522 (2002);
  %[arXiv:hep-th/0112249].
  %%CITATION = PHRVA,D65,103522;%%
%
%\cite{Cai:2008qw}
%\bibitem{Cai:2008qw}
  Y.~F.~Cai, T.~t.~Qiu, R.~Brandenberger and X.~m.~Zhang,
  %``A Nonsingular Cosmology with a Scale-Invariant Spectrum of Cosmological
  %Perturbations from Lee-Wick Theory,''
  Phys.\ Rev.\  D {\bf 80}, 023511 (2009).
  %[arXiv:0810.4677 [hep-th]].
  %%CITATION = PHRVA,D80,023511;%%

%\cite{Starobinsky:1979ty}
\bibitem{Starobinsky:1979ty}
  see also
  A.~A.~Starobinsky,
  %``Spectrum of relict gravitational radiation and the early state of the
  %universe,''
  JETP Lett.\  {\bf 30} (1979) 682.
  %[Pisma Zh.\ Eksp.\ Teor.\ Fiz.\  {\bf 30} (1979) 719].
  %%CITATION = ZFPRA,30,719;%%

%====================non-stand INf=====

%\cite{Peiris:2007gz}
\bibitem{Peiris:2007gz}
  H.~V.~Peiris, D.~Baumann, B.~Friedman and A.~Cooray,
  %``Phenomenology of D-Brane Inflation with General Speed of Sound,''
  Phys.\ Rev.\  D {\bf 76}, 103517 (2007);
  %[arXiv:0706.1240 [astro-ph]].
  %%CITATION = PHRVA,D76,103517;%%
%
%\cite{Bean:2007eh}
%\bibitem{Bean:2007eh}
  R.~Bean, X.~Chen, H.~Peiris and J.~Xu,
  %``Comparing Infrared Dirac-Born-Infeld Brane Inflation to Observations,''
  Phys.\ Rev.\  D {\bf 77}, 023527 (2008).
  %[arXiv:0710.1812 [hep-th]].
  %%CITATION = PHRVA,D77,023527;%%

%\cite{Hall:2007qw}
\bibitem{Hall:2007qw}
  L.~M.~H.~Hall and H.~V.~Peiris,
  %``Cosmological Constraints on Dissipative Models of Inflation,''
  JCAP {\bf 0801}, 027 (2008).
  %[arXiv:0709.2912 [astro-ph]].
  %%CITATION = JCAPA,0801,027;%%

%\cite{Easther:2004vq}
\bibitem{Easther:2004vq}
  R.~Easther, W.~H.~Kinney and H.~Peiris,
  %``Observing trans-Planckian signatures in the cosmic microwave  background,''
  JCAP {\bf 0505}, 009 (2005).
  %[arXiv:astro-ph/0412613].
  %%CITATION = JCAPA,0505,009;%%

%\cite{Spergel:2006hy}
\bibitem{Spergel:2006hy}
  see also Section 5 of:
  D.~N.~Spergel {\it et al.}  [WMAP Collaboration],
  %``Wilkinson Microwave Anisotropy Probe (WMAP) three year results:
  %Implications for cosmology,''
  Astrophys.\ J.\ Suppl.\  {\bf 170}, 377 (2007).
  %[arXiv:astro-ph/0603449].
  %%CITATION = APJSA,170,377;%%

%================cosmological perturbations==============

%\cite{Mukhanov:1990me}
\bibitem{Mukhanov:1990me}
  V.~F.~Mukhanov, H.~A.~Feldman and R.~H.~Brandenberger,
  %``Theory of cosmological perturbations. Part 1. Classical perturbations. Part
  %2. Quantum theory of perturbations. Part 3. Extensions,''
  Phys.\ Rept.\  {\bf 215}, 203 (1992).
  %%CITATION = PRPLC,215,203;%%

%=======================Data=====================

%\cite{Komatsu:2010fb}
\bibitem{Komatsu:2010fb}
  E.~Komatsu {\it et al.},
  %``Seven-Year Wilkinson Microwave Anisotropy Probe (WMAP) Observations:
  %Cosmological Interpretation,''
  arXiv:1001.4538 [astro-ph.CO].
  %%CITATION = ARXIV:1001.4538;%%

%\cite{Tegmark:2003uf}
\bibitem{Tegmark:2003uf}
%  M.~Tegmark {\it et al.},  %[SDSS Collaboration],
  %``The 3D power spectrum of galaxies from the SDSS,''
%  Astrophys.\ J.\  {\bf 606}, 702 (2004);
  %[arXiv:astro-ph/0310725].
  %%CITATION = ASJOA,606,702;%%
%
%\cite{Tegmark:2006az}
%\bibitem{Tegmark:2006az}
  M.~Tegmark {\it et al.},  %[SDSS Collaboration],
  %``Cosmological Constraints from the SDSS Luminous Red Galaxies,''
  Phys.\ Rev.\  D {\bf 74}, 123507 (2006).
  %[arXiv:astro-ph/0608632].
  %%CITATION = PHRVA,D74,123507;%%


%=========Global Fitting============

%\cite{Lewis:2002ah}
\bibitem{Lewis:2002ah}
  A.~Lewis and S.~Bridle,
  %``Cosmological parameters from CMB and other data: a Monte-Carlo approach,''
  Phys.\ Rev.\  D {\bf 66}, 103511 (2002);
  %[arXiv:astro-ph/0205436];
  %%CITATION = PHRVA,D66,103511;%%
  Available from
  http://cosmologist.info/cosmomc.

%\cite{Kowalski:2008ez}
\bibitem{Kowalski:2008ez}
  M.~Kowalski {\it et al.},  %[Supernova Cosmology Project Collaboration],
  %``Improved Cosmological Constraints from New, Old and Combined Supernova
  %Datasets,''
  Astrophys.\ J.\  {\bf 686}, 749 (2008).
  %[arXiv:0804.4142 [astro-ph]].
  %%CITATION = ASJOA,686,749;%%

%\cite{DiPietro:2002cz}
\bibitem{DiPietro:2002cz}
  E.~Di Pietro and J.~F.~Claeskens,
  %``Quintessence models faced with future supernovae data,''
  Mon.\ Not.\ Roy.\ Astron.\ Soc.\  {\bf 341}, 1299 (2003).
  %[arXiv:astro-ph/0207332].
  %%CITATION = MNRAA,341,1299;%%

%\cite{Riess:2009pu}
\bibitem{Riess:2009pu}
  A.~G.~Riess {\it et al.},
  %``A Redetermination of the Hubble Constant with the Hubble Space Telescope
  %from a Differential Distance Ladder,''
  Astrophys.\ J.\  {\bf 699}, 539 (2009).
  %[arXiv:0905.0695 [astro-ph.CO]].
  %%CITATION = ASJOA,699,539;%%

%==============bounce perturbation===========

%\cite{Cai:2009fn}
%\bibitem{Cai:2009fn}
  %Y.~F.~Cai, W.~Xue, R.~Brandenberger and X.~Zhang,
  %``Non-Gaussianity in a Matter Bounce,''
  %JCAP {\bf 0905}, 011 (2009);
  %[arXiv:0903.0631 [astro-ph.CO]].
  %%CITATION = JCAPA,0905,011;%%
%
%\cite{Cai:2009rd}
%\bibitem{Cai:2009rd}
  %Y.~F.~Cai, W.~Xue, R.~Brandenberger and X.~m.~Zhang,
  %``Thermal Fluctuations and Bouncing Cosmologies,''
  %JCAP {\bf 0906}, 037 (2009);
  %[arXiv:0903.4938 [hep-th]].
  %%CITATION = JCAPA,0906,037;%%
%
%\cite{Cai:2009hc}
%\bibitem{Cai:2009hc}
  %Y.~F.~Cai and X.~Zhang,
  %``Primordial perturbation with a modified dispersion relation,''
  %Phys.\ Rev.\  D {\bf 80}, 043520 (2009).
  %[arXiv:0906.3341 [astro-ph.CO]].
  %%CITATION = PHRVA,D80,043520;%%

%==============Other bounce===================

%\cite{Gasperini:1992em}
\bibitem{Gasperini:1992em}
  M.~Gasperini and G.~Veneziano,
  %``Pre - big bang in string cosmology,''
  Astropart.\ Phys.\  {\bf 1}, 317 (1993).
  %[arXiv:hep-th/9211021].
  %%CITATION = APHYE,1,317;%%

%\cite{Khoury:2001wf}
\bibitem{Khoury:2001wf}
  J.~Khoury, B.~A.~Ovrut, P.~J.~Steinhardt and N.~Turok,
  %``The ekpyrotic universe: Colliding branes and the origin of the hot big
  %bang,''
  Phys.\ Rev.\  D {\bf 64}, 123522 (2001).
  %[arXiv:hep-th/0103239].
  %%CITATION = PHRVA,D64,123522;%%

%\cite{Brustein:1994kn}
\bibitem{Brustein:1994kn}
  R.~Brustein {\it et al.},
  %M.~Gasperini, M.~Giovannini, V.~F.~Mukhanov and G.~Veneziano,
  %``Metric perturbations in dilaton driven inflation,''
  Phys.\ Rev.\  D {\bf 51}, 6744 (1995).
  %[arXiv:hep-th/9501066].
  %%CITATION = PHRVA,D51,6744;%%

%\cite{Lyth:2001pf}
\bibitem{Lyth:2001pf}
  D.~H.~Lyth,
  %``The primordial curvature perturbation in the ekpyrotic universe,''
  Phys.\ Lett.\  B {\bf 524}, 1 (2002).
  %[arXiv:hep-ph/0106153].
  %%CITATION = PHLTA,B524,1;%%

%\cite{Mielczarek:2010ga}
\bibitem{Mielczarek:2010ga}
  J.~Mielczarek, M.~Kamionka, A.~Kurek and M.~Szydlowski,
  %``Observational hints on the Big Bounce,''
  JCAP {\bf 1007}, 004 (2010).
  %[arXiv:1005.0814 [gr-qc]].
  %%CITATION = JCAPA,1007,004;%%



\end{thebibliography}
\end{document}